\documentclass[9pt]{article}
\begin{document}
\pagestyle{plain}
\setcounter{page}{1}
\begin{center}
{\large\bf Quantum Gravity Resolution to the Cosmological Constant Problem}
\vskip 0.3 true in
{\large J. W. Moffat}
\vskip 0.3 true in
{\it Department of Physics, University of Toronto,
Toronto, Ontario M5S 1A7, Canada}
\end{center}
\begin{abstract}%
A finite quantum gravity theory is used to resolve the cosmological constant
problem. A fundamental quantum gravity scale, $\Lambda_G\leq
10^{-3}$ eV, is introduced above which the quantum corrections to the vacuum energy
density coupled to gravity are exponentially suppressed by a graviton vertex form
factor, yielding an observationally acceptable value for the particle physics
contribution to the cosmological constant. Classical Einstein gravity retains its
causal behavior as well as the standard agreement with observational data.
\end{abstract}



\section{Introduction}

The cosmological constant problem is considered to be the most
severe hierarchy problem in modern
physics~\cite{Weinberg2,Sahni,Witten2,Carroll,Straumann,Rugh}.
We shall propose a quantum gravity solution to the problem,
based on a nonlocal, finite quantum field theory and quantum gravity
theory~\cite{Moffat,Moffat2,Moffat3,Woodard,Kleppe2,Cornish,Cornish2,
Hand,Woodard2,Clayton,Paris,Troost,Joglekar,Joglekar2,Joglekar3,Moffat4}.

We can define an effective cosmological constant
\begin{equation}
\lambda_{\rm eff}=\lambda_0+\lambda_{\rm vac},
\end{equation}
where $\lambda_0$ is the `bare' cosmological
constant in Einstein's classical field equations,
and $\lambda_{\rm vac}$ is the contribution that arises from the
vacuum density $\lambda_{\rm vac}=8\pi G\rho_{\rm vac}$. Already
at the standard model electroweak scale $\sim 10^2$ GeV, a
calculation of the vacuum density $\rho_{\rm vac}$, based on local quantum field
theory results in a discrepancy with the observational bound
\begin{equation}
\label{vacbound}
\rho_{\rm vac} \leq 10^{-47}\, ({\rm GeV})^4,
\end{equation}
of order $10^{55}$, resulting in a a severe fine tuning problem, since the virtual
quantum fluctuations giving rise to $\lambda_{\rm vac}$ must cancel
$\lambda_0$ to an unbelievable degree of accuracy. The bound on $\lambda_{\rm vac}$
is
\begin{equation}
\label{lambdabound}
\lambda_{\rm vac} \leq 10^{-84}\,{\rm GeV}^2.
\end{equation}
If we choose the quantum gravity scale $\Lambda_G\leq 10^{-3}$ eV,
then our nonlocal quantum gravity theory leads to a damping of the
gravitational quantum corrections to $\lambda_0$ for $q^2 \gg \Lambda_G^2$, where $q$
is the Euclidean internal loop momentum. This suppresses
$\lambda_{\rm vac}$ below the observational bound (\ref{lambdabound}). Since the
graviton tree graphs are identical to the standard point like, local tree graphs of
perturbative gravity, we retain classical, causal GR and Newtonian gravity theory,
and the measured value of the gravitational constant $G$. Only the quantum gravity
loop graphs are suppressed above energies $\leq 10^{-3}$ eV.

The scales $\Lambda_{SM}$ and $\Lambda_G$ are determined by the quantum
non-localizable nature of the standard model (SM) particles as compared to the
graviton. The SM particle radiative corrections have a nonlocal scale
at $\Lambda_{SM} > 1-10$ TeV or a length scale $\ell_{SM} < 10^{-16}$ cm, whereas the
graviton radiative corrections are localizable down to an energy scale
$\Lambda_G\leq 10^{-3}$ eV or a length scale $\ell_G < 1$ cm. Thus, the fundamental
energy scales in the theory are determined by the underlying physical nature of the
particles and fields and do not correspond to arbitrary cut-offs, which destroy the
gauge invariances of the field theory. The underlying explanation of these physical
scales must be sought in a more fundamental theory.

The `fuzziness' of the SM particles and the graviton, due to the nonlocal nature of
the quantum field theory, gives rise to the `composite' nature of the particles. An
attempt to incorporate a composite graviton in a toy model field theory was made by
Sundrum~\cite{Sundrum}. In this model, the `stringy' graviton was coupled to a
stringy halo surrounding the SM particles in the loop coupled to external gravitons.

In Section 2, we describe the local action of the theory
and in Section 3, we provide a review of the basic properties of the finite quantum
field theory as a perturbative scheme.
In Section 4, we develop the formalism for quantum gravity, while in Section 5, we
analyze the results of gluon and gravitational vacuum polarization calculations. In
Section 6, we use the quantum gravity theory to resolve the cosmological constant
problem and in Section 7, we end with concluding remarks.

\section{\bf The Action}

We begin with the four-dimensional action
\begin{equation}
W=W_{\rm grav}+W_{YM}+W_{\rm H}+W_{\rm Dirac}+W_M,
\end{equation}
where
\begin{equation}
\label{EinsteinHilbert}
W_{\rm grav}=-\frac{2}{\kappa^2}\int d^4x\sqrt{-g}(R+2\lambda_0),
\end{equation}
\begin{equation}
W_{\rm YM}=-\frac{1}{4}\int d^4x\sqrt{-g}\,{\rm Tr}(F^2),
\end{equation}
\begin{equation}
W_{\rm H}=-\int
d^4x\sqrt{-g}\biggl[\frac{1}{2}D_\mu\phi^iD^\mu\phi^i+V(\phi^2)\biggr],
\end{equation}
\begin{equation}
W_{\rm Dirac}=\frac{1}{2}\int
d^4x\sqrt{-g}\bar{\psi}\gamma^ae_a^\mu[\partial_\mu\psi-\omega_\mu\psi
-{\cal D}(A_{i\mu})\psi]+h.c.
\end{equation}
Here, we use the
notation: $\mu,\nu=0,1,2,3$, $g={\rm det}(g_{\mu\nu})$ and the
metric signature of Minkowski spacetime is $\eta_{\mu\nu}={\rm
diag}(-1,+1,+1,+1)$. The Riemann tensor is defined such that
\begin{equation}
{R^\lambda}_{\mu\nu\rho}=\partial_\rho{\Gamma_{\mu\nu}}^\lambda
-\partial_\nu{\Gamma_{\mu\rho}}^\lambda+
{\Gamma_{\mu\nu}}^\alpha{\Gamma_{\rho\alpha}}^\lambda
-{\Gamma_{\mu\rho}}^\alpha{\Gamma_{\nu\alpha}}^\lambda.
\end{equation}
Moreover, h.c. denotes the Hermitian conjugate,
$\bar\psi=\psi^{\dagger} \gamma^0$, and $e^\mu_a$ is a vierbein,
related to the metric by
\begin{equation}
g_{\mu\nu}=\eta_{ab}e_\mu^ae_\nu^b,
\end{equation} where
$\eta_{ab}$ is the four-dimensional Minkowski metric tensor
associated with the flat tangent space with indices a,b,c...
Moreover, $F^2= F^i_{\mu\nu}F^{i\mu\nu}$, $R$ denotes the scalar
curvature and
\begin{equation} F_{i\mu\nu}=\partial_\nu A_{i\mu}-\partial_\mu
A_{i\nu}-ef_{ikl}A_{k\mu}A_{l\nu},
\end{equation} where
$A_{i\mu}$ are the gauge fields of the Yang-Mills group with
generators $f_{ikl}$, $e$ is the coupling constant and
$\kappa^2=32\pi G$ with $c=1$. We denote by $D_\mu$ the
covariant derivative operator
\begin{equation}
D_\mu\phi^i=\partial_\mu\phi^i+ef^{ikl}A_\mu^k\phi^l.
\end{equation}
The Higgs potential $V(\phi^2)$ is of the form
leading to spontaneous symmetry breaking
\begin{equation}
V(\phi^2)=\frac{1}{4}g(\phi^i\phi^i-K^2)^2+V_0,
\end{equation}
where $V_0$ is an adjustable constant and the coupling constant
$g > 0$.

The spinor field is minimally coupled to the gauge potential
$A_{i\mu}$, and ${\cal D}$ is a matrix representation of the
gauge group $SO(3,1)$. The spin connection $\omega_\mu$ is
\begin{equation}
\omega_\mu=\frac{1}{2}\omega_{\mu ab}\Sigma^{ab},
\end{equation} where
$\Sigma^{ab}=\frac{1}{4}[\gamma^a,\gamma^b]$ is the spinor matrix
associated with the Lorentz algebra $SO(3,1)$. The components
$\omega_{\mu ab}$ satisfy
\begin{equation}
\partial_\mu
e^\sigma_a+{\Gamma_{\mu\nu}}^\sigma e^\nu_a-{\omega_{\mu a}}^\rho
e_\rho^\sigma=0,
\end{equation} where ${\Gamma_{\mu\nu}}^\sigma$
is the Christoffel symbol. The field equations for the
gravity-Yang-Mills-Higgs-Dirac sector are
\begin{equation}
R_{\mu\nu}-\frac{1}{2}g_{\mu\nu}R-\lambda
g_{\mu\nu}=-\frac{1}{4}\kappa^2T_{\mu\nu},
\end{equation}
\begin{equation}
g^{\rho\mu}\nabla_\rho
F^i_{\mu\nu}=g^{\rho\mu}\biggl(\partial_\rho
F^i_{\mu\nu}-{\Gamma_{\rho\mu}}^\sigma F^i_{\sigma\nu}
-{\Gamma_{\rho\nu}}^\sigma F^i_{\mu\sigma}
+[A_\rho,F_{\mu\nu}]^i\biggr)=0,
\end{equation}
\begin{equation}
\frac{1}{\sqrt{-g}}D_\mu[\sqrt{-g}g^{\mu\nu}D_\nu\phi^i]
=\biggl(\frac{\partial V}{\partial\phi^2}\biggr)\phi^i,
\end{equation}
\begin{equation}
\gamma^ae^\mu_a[\partial_\mu-\omega_\mu-{\cal D}(A_\mu)]\psi=0.
\end{equation}
The energy-momentum tensor is
\begin{equation}
T_{\mu\nu}=T^{\rm YMH}_{\mu\nu}+T^{\rm Dirac}_{\mu\nu}+T^{\rm
M}_{\mu\nu},
\end{equation}
where
\begin{equation}
T_{\mu\nu}^{\rm YMH}=F^i_{\mu\sigma}
F^{i\sigma}_\nu+D_\mu\phi^iD_\nu\phi^i
$$ $$
-\frac{1}{2}g_{\mu\nu}\biggl[\frac{1}{2}{\rm Tr}(F^2)
+D_\sigma\phi^iD^\sigma\phi^i+V(\phi^2)\biggr],
\end{equation}
\begin{equation}
T^{\rm Dirac}_{\mu\nu}
=-\bar\psi\gamma_\mu[\partial_\nu-\omega_\nu
-{\cal D}(A_{i\nu})]\psi,
\end{equation}
and $T_{\mu\nu}^{\rm M}$ is the energy-momentum tensor of
non-field matter.

\section{\bf Finite Quantum Field Theory Formalism}

An important development in nonlocal quantum field theory was the discovery that
gauge invariance and unitarity can be restored by adding series
of higher interactions. The resulting theory possesses a
nonlinear, field representation dependent gauge invariance which
agrees with the original local symmetry on-shell but is larger
off-shell. Quantization is performed in the functional formalism
using an analytic and convergent measure factor which retains
invariance under the new symmetry. An explicit calculation was
made of the measure factor in QED~\cite{Moffat2}, and it
was obtained to lowest order in Yang-Mills
theory~\cite{Kleppe2}. Kleppe and
Woodard~\cite{Woodard2} obtained an ansatz based on the
derived dimensionally regulated result when
$\Lambda\rightarrow\infty$, which was conjectured to lead to a
general functional measure factor in nonlocal gauge theories.

In contrast to string theory, we can achieve a {\it genuine
quantum field theory}, which allows vertex operators to be taken off the
mass shell. The finiteness draws from the fact that factors of $\exp[{\cal
K}(p^2)/2\Lambda^2]$ are attached to propagators which suppress any
ultraviolet divergences in Euclidean momentum space, where $\Lambda$ is an
energy scale factor. An important feature of the field theory is {\it that only the
quantum loop graphs have nonlocal properties}; the classical tree graph
theory retains full causal and local behavior.
 
We shall consider
the 4-dimensional spacetime to be approximately flat Minkowski
spacetime. Let us denote by $f$ a generic local field and write
the standard local Lagrangian as
\begin{equation}
{\cal L}[f]={\cal L}_F[f]+{\cal L}_I[f],
\end{equation}
where ${\cal L}_F$ and ${\cal L}_I$ denote the free part and the interaction part
of the action, respectively, and
\begin{equation}
{\cal L}_F[f]=\frac{1}{2}f_i{\cal K}_{ij}f_j.
\end{equation}
In a gauge theory the action would be the Becchi, Rouet, Stora, Tyutin (BRST)
gauge-fixed action including ghost fields in the invariant action required
to fix the gauge\cite{Becchi}. The kinetic operator ${\cal K}$ is fixed by
defining a Lorentz-invariant distribution operator
\begin{equation}
\label{distribution}
{\cal E}\equiv \exp\biggl(\frac{{\cal K}}{2\Lambda^2}\biggr)
\end{equation} and the operator:
\begin{equation}
{\cal O}=\frac{{\cal E}^2-1}{{\cal
K}}=\int_0^1\frac{d\tau}{\Lambda^2}\exp\biggl(\tau\frac{{\cal K}}{\Lambda^2}\biggr).
\end{equation}

The regularized interaction Lagrangian takes the form
\begin{equation}
{\hat {\cal L}}_{I}=-\sum_n(-g)^nf{\cal I}[{\cal F}^n,{\cal O}^{(n-1)})]f,
\end{equation}
where $g$ is a coupling constant and ${\cal F}$ is a vertex function form factor. The
decomposition of ${\cal I}$ in order $n=2$ is such that the operator ${\cal O}$
splits into two parts ${\cal F}^2/{\cal K}$ and $-1/{\cal K}$. For Compton amplitudes
the first such term cancels the contribution from the corresponding lower order
channel, while the second term is just the usual local field theory result for that
channel. The action is then invariant under an extended nonlocal gauge
transformation. The precise results for QED were described in ref.~\cite{Moffat2}.

The regularized action is found by expanding
${\hat{\cal L}}_I$ in an infinite series of interaction terms. Since ${\cal F}$ and
${\cal O}$ are entire function of ${\cal K}$ the higher interactions are also
entire functions of ${\cal K}$. This is important for preserving the Cutkosky rules
and unitarity, for an entire function does not possess any singularities in the
finite complex momentum plane.

The regulated action is gauge invariant under the transformation
\begin{equation}
\delta f=ig\int d^4yd^4z{\cal T}[f](x,y,z)\theta(y)f,
\end{equation}
where $\theta$ is the infinitesimal gauge parameter and ${\cal T}$ is a
spinorial matrix for $f(x)=\psi(x)$ as well as a function of a gauge potential.
An explicit construction for QED~\cite{Moffat2} was given using the Cutkosky rules as
applied to the nonlocal field theory, whose propagators have poles only where ${\cal
K}=0$ and whose vertices are entire functions of ${\cal K}$. The regulated action
satisfies these requirements which guarantees unitarity on the physical space of
states. The local action is gauge fixed and then a regularization is performed on
the BRST theory.

Quantization is performed using the definition
\begin{equation}
\langle 0\vert T^*(O[f])\vert 0\rangle_{\cal E}=\int[Df]\mu[f]({\rm gauge\,
fixing})O[f]\exp(i\hat W[f]).
\end{equation}
where ${\hat W}$ is the regulated action. On the left-hand side we have the regulated
vacuum expectation value of the $T^*$-ordered product of an arbitrary operator
$O[f]$ formed from the local fields $f$. The subscript ${\cal E}$ signifies that a
regulating Lorentz distribution has been used. Moreover, $\mu[f]$ is a gauge
invariant measure factor and there is a gauge fixing factor, both of which are needed
to maintain perturbative unitarity in gauge theories.

The new Feynman rules for are obtained as follows: Every leg of a
diagram is connected to a local propagator,
\begin{equation}
\label{regpropagator}
D(q^2)=\frac{i}{{\cal K}(q^2)+i\epsilon}
\end{equation}
and every vertex has a form factor ${\cal F}^k(q^2)$, where $q$ is the momentum
attached to the propagator $D(q^2)$, which has the form
\begin{equation}
{\cal F}^k(q^2)\equiv{\cal E}^k(q^2)=\exp\biggl(\frac{\cal K}{2\Lambda_k}\biggr),
\end{equation}
where $k$ denotes the particle nature of the external leg in the vertex.
The formalism is set up in Minkowski spacetime and loop integrals are formally
defined in Euclidean space by performing a Wick rotation. This facilitates the
analytic continuation; the whole formalism could from the outset be developed in
Euclidean space.

Renormalization is carried out as in any other field theory.
The bare parameters are calculated from the renormalized ones and
$\Lambda$, such that the limit $\Lambda\rightarrow\infty$ is finite for
all noncoincident Green's functions, and the bare parameters are those of
the local theory. The regularizing interactions {\it are determined by the
local operators.}

The regulating Lorentz distribution function ${\cal E}$ must be chosen to
perform an explicit calculation in perturbation theory. We do not know the
unique choice of ${\cal E}$. However, once a choice for the function is made, then the
theory and the perturbative calculations are uniquely fixed. A standard
choice in early papers is~\cite{Moffat,Moffat2}:
\begin{equation}
\label{reg}
{\cal E}_m=\exp\biggl(\frac{\partial^2-m^2}{2\Lambda^2}\biggr).
\end{equation}

In the tree order, Green's functions remain local except for
external lines which are unity on-shell. It follows immediately that since
on-shell tree amplitudes are unchanged by the regularization, the Lagrangian
preserves all symmetries on-shell. Also all loops contain at least
one regularizing  vertex function and therefore are ultraviolet finite.

The on-shell tree amplitudes agree with the local,
unregulated action, while the loop amplitudes disagree. This
seems to contradict the Feynman tree
theorem~\cite{Feynman3}, which states that loop
amplitudes of local field theory can be expressed as sums of
integrals of tree diagrams. If two local theories agree at
the tree level, then the loop amplitudes agree as well.
However, the tree theorem does not apply to nonlocal field
theories. The tree theorem is proved by using the propagator relation
\begin{equation}
D_F=D_R+D^+
\end{equation}
to expand the Feynman propagator $D_F$ into a series in the on-
shell propagator $D^+$. This decomposes all terms with even one
$D^+$ into trees. The term with no $D^+s$ is a loop formed with
the retarded propagator and vanishes for local interactions. But
for nonlocal interactions, this term generally survives and new
physical effects occur in loop amplitudes, which cannot be
predicted from the local on-shell tree graphs.

\section{Finite Perturbative Quantum Gravity}

As is well-known, the problem with perturbative quantum gravity
based on a point-like graviton and a local field theory
formalism is that the theory is not
renormalizable~\cite{Veltman,Van}. Due to the Gauss-Bonnet
theorem, it can be shown that the one-loop graviton calculation
is renormalizable but two-loop is not~\cite{Sagnotti}.
Moreover, gravity-matter interactions are not renormalizable at
any loop order.

We shall now formulate the gravitational sector
in more detail. This problem has been considered
previously in the context of four-dimensional
GR~\cite{Moffat,Moffat2,Moffat4}. We expand the
gravity sector about flat Minkowski spacetime. In fact, our quantum gravity theory
can be formulated as a perturbative theory by expanding around any
fixed, classical metric background~\cite{Veltman}
\begin{equation}
\label{background}
g_{\mu\nu}={\bar g}_{\mu\nu}+h_{\mu\nu},
\end{equation}
where ${\bar g}_{\mu\nu}$ is any smooth background metric field,
e.g. a de Sitter spacetime metric. For the sake of simplicity, we
shall only consider expansions about flat spacetime. Since the
gravitational field is weak up to the Planck energy scale, this
expansion is considered justified; even at the standard model
energy scale $E_{\rm SM}\sim 10^2$ GeV,
the curvature of spacetime is very small. However, if we wish to include the
cosmological constant, then we cannot strictly speaking
expand about flat spacetime. This is to be expected, because the cosmological
constant produces a curved spacetime even when the
energy-momentum tensor $T_{\mu\nu}=0$. Therefore, we should
use the expansion (\ref{background}) with a curved background metric. But for
energy scales encountered in particle physics, the curvature is very small, so we can
approximate the perturbation caculation by using the flat spacetime expansion and
trust that the results are valid in general for curved spacetime backgrounds
including the cosmological constant.

Let us define ${\bf g}^{\mu\nu}=\sqrt{-g}g^{\mu\nu}$.
It can be shown that $\sqrt{-g}=\sqrt{-{\bf g}}$, where ${\bf g}=
{\rm det}({\bf g}^{\mu\nu})$ and $\partial_\rho{\bf g}={\bf
g}_{\alpha\beta}\partial_\rho{\bf g}^{\alpha\beta}{\bf g}$. We
can then write the local gravitational action $W_{\rm grav}$ in
the form~\cite{Goldberg}:
\begin{equation}
\label{action}
W_{\rm grav}=\int d^4x{\cal L}_{\rm grav}=\frac{1}{2\kappa^2}\int
d^4x [({\bf g}^{\rho\sigma}{\bf g}_{\lambda\mu} {\bf
g}_{\kappa\nu}
$$ $$
-\frac{1}{2}{\bf g}^{\rho\sigma} {\bf
g}_{\mu\kappa}{\bf g}_{\lambda\nu}
-2\delta^\sigma_\kappa\delta^\rho_\lambda{\bf
g}_{\mu\nu})\partial_\rho{\bf g}^{\mu\kappa} \partial_\sigma{\bf
g}^{\lambda\nu}
$$ $$
-\frac{1}{\alpha\kappa^2}\partial_\mu{\bf
g}^{\mu\nu}\partial_\kappa{\bf g}^{\kappa\lambda}
\eta_{\nu\lambda}
+{\bar C}^\nu\partial^\mu X_{\mu\nu\lambda}C^\lambda],
\end{equation}
where we have added a gauge fixing term with the parameter
$\alpha$, $C^\mu$ is the Fadeev-Popov ghost field and
$X_{\mu\nu\lambda}$ is a differential operator.

We expand the local interpolating graviton field ${\bf
g}^{\mu\nu}$ as
\begin{equation}
{\bf g}^{\mu\nu}=\eta^{\mu\nu}+\kappa\gamma^{\mu\nu}+O(\kappa^2).
\end{equation} Then,
\begin{equation}
{\bf g}_{\mu\nu}=\eta_{\mu\nu}-\kappa\gamma_{\mu\nu}
+\kappa^2{\gamma_\mu}^\alpha{\gamma_\alpha}_\nu+O(\kappa^3).
\end{equation}

The gravitational Lagrangian density is expanded as
\begin{equation}
{\cal L}_{\rm grav}={\cal L}^{(0)}+\kappa{\cal L}^{(1)}
+\kappa^2{\cal L}^{(2)}+....
\end{equation}
We obtain
\begin{equation}
{\cal L}^{(0)}=\frac{1}{2}\partial_\sigma\gamma_{\lambda\rho}
\partial^\sigma\gamma^{\lambda\rho}
-\partial_\lambda\gamma^{\rho\kappa}
\partial_\kappa\gamma^\lambda_\rho
-\frac{1}{4}\partial_\rho\partial^\rho\gamma
$$ $$
-\frac{1}{\alpha}\partial_\rho\gamma^\rho_\lambda\partial_\kappa
\gamma^{\kappa\lambda}
+{\bar C}^\lambda\partial_\sigma\partial^\sigma C_\lambda,
\end{equation}
\begin{equation}
{\cal L}^{(1)}
=\frac{1}{4}(-4\gamma_{\lambda\mu}\partial^\rho\gamma^{\mu\kappa}
\partial_\rho\gamma^\lambda_\kappa+2\gamma_{\mu\kappa}
\partial^\rho\gamma^{\mu\kappa}\partial_\rho\gamma
$$ $$
+2\gamma^{\rho\sigma}\partial_\rho\gamma_{\lambda\nu}
\partial_\sigma\gamma^{\lambda\nu}
-\gamma^{\rho\sigma}\partial_\rho\gamma\partial_\sigma\gamma
+4\gamma_{\mu\nu}\partial_\lambda\gamma^{\mu\kappa}
\partial_\kappa\gamma^{\nu\lambda})
$$ $$
+{\bar C}^\nu\gamma_{\kappa\mu}\partial^\kappa\partial^\mu C_\nu
+{\bar C}^\nu\partial^\mu\gamma_{\kappa\mu}\partial^\kappa C_\nu
-{\bar C}^\nu\partial^\lambda\partial^\mu\gamma_{\mu\nu}C_\lambda
-{\bar C}^\nu\partial^\mu\gamma_{\mu\nu}\partial^\lambda
C_\lambda,
\end{equation}
\begin{equation}
{\cal L}^{(2)}=\frac{1}{4}(4\gamma_{\kappa\alpha}
\gamma^{\alpha\nu}
\partial^\rho\gamma^{\lambda\kappa}\partial_\rho\gamma_{\nu\lambda}
+(2\gamma_{\lambda\mu}\gamma_{\kappa\nu}-\gamma_{\mu\kappa}\gamma_{\nu\lambda})
\partial^\rho\gamma^{\mu\kappa}\partial_\rho\gamma^{\nu\lambda}
$$ $$
-2\gamma_{\lambda\alpha}\gamma^\alpha_\nu\partial^\rho\gamma^{\lambda\nu}
\partial_\rho\gamma-2\gamma^{\rho\sigma}\gamma^\kappa_\nu
\partial_\rho\gamma_{\lambda\kappa}\partial_\sigma\gamma^{\nu\lambda}
$$ $$
+\gamma^{\rho\sigma}\gamma^{\nu\lambda}\partial_\sigma\gamma_{\nu\lambda}
\partial_\rho\gamma-2\gamma_{\mu\alpha}\gamma^{\alpha\nu}
\partial^\lambda\gamma^{\mu\kappa}\partial_\kappa\gamma_{\nu\lambda}),
\end{equation}
where $\gamma={\gamma^\alpha}_\alpha$.

In the limit $\alpha\rightarrow\infty$, the Lagrangian density
${\cal L}_{\rm grav}$ is invariant under the gauge
transformation
\begin{equation}
\delta\gamma_{\mu\nu}=X_{\mu\nu\lambda}\xi^\lambda,
\end{equation}
where $\xi^\lambda$ is an infinitesimal vector quantity and
\begin{equation}
X_{\mu\nu\lambda}=\kappa(-\partial_\lambda\gamma_{\mu\nu}
+2\eta_{(\mu\lambda}\gamma_{\kappa\nu)}\partial^\kappa)
+(\eta_{(\mu\lambda}\partial_{\nu)}-\eta_{\mu\nu}\partial_\lambda).
\end{equation}
However, for the quantized theory it is more useful to require
the BRST symmetry. We choose $\xi^\lambda=C^\lambda\sigma$,
where $\sigma$ is a global anticommuting scalar. Then, the BRST
transformation is
\begin{equation}
\delta\gamma_{\mu\nu}=X_{\mu\nu\lambda}C^\lambda\sigma,
\quad \delta {\bar C}^\nu=-\partial_\mu\gamma^{\mu\nu}
\biggl(\frac{2\sigma}{\alpha}\biggr),\quad \delta C_\nu=\kappa
C^\mu \partial_\mu C_\nu\sigma.
\end{equation}

We now substitute the operators
\begin{equation}
\gamma_{\mu\nu}\rightarrow{\hat\gamma}_{\mu\nu},\quad
C_\lambda\rightarrow {\hat C}_\lambda,\quad
{\bar C}_\nu\rightarrow {\hat{\bar C}}_\nu,
\end{equation}
where
\begin{equation}
\label{hatgamma}
{\hat \gamma}_{\mu\nu}={\cal E}^{-1}\gamma_{\mu\nu},
\quad {\hat C}_\lambda={\cal E}^{-1}C_\lambda,\quad
{\hat{\bar C}}_\lambda={\cal E}^{-1}C_\lambda.
\end{equation}

The regularized
Lagrangian density up to order $\kappa^2$ is invariant under the
extended BRST transformations~\cite{Cornish}:
\begin{equation}
{\hat\delta}_0\gamma_{\mu\nu}=X^{(0)}_{\mu\nu\lambda}
C^\lambda\sigma=(\partial_\nu C_\mu+\partial_\mu
C_\nu-\eta_{\mu\nu}\partial_\lambda C^\lambda)\sigma,
\end{equation}
\begin{equation}
{\hat\delta}_1\gamma_{\mu\nu}=\kappa{\cal
E}_0^2X^{(1)}_{\mu\nu\lambda} C^\lambda_\sigma
=\kappa{\cal E}_0^2(2\gamma_{\rho(\mu}\partial^\rho C_{\nu)}
-\partial_\lambda\gamma_{\mu\nu}
C^\lambda-\gamma_{\mu\nu}\partial_\lambda C^\lambda)\sigma,
\end{equation}
\begin{equation}
{\hat\delta}_0{\bar C}^\nu=2\partial_\mu\gamma^{\mu\nu}\sigma,
\end{equation}
\begin{equation}
{\hat\delta}_1C_\nu=\kappa{\cal E}_0^2C^\mu \partial_\mu
C_\nu\sigma.
\end{equation}

The order $\kappa^2$ transformations are
\begin{equation}
{\hat\delta}_2\gamma_{\mu\nu}=\kappa^2{\cal E}_0^2[2\partial^\rho
C_{(\nu}
{\tilde D}_{\mu)\rho\kappa\lambda}
(B^{\kappa\lambda}+H^{\kappa\lambda})
$$ $$
-C^\rho{\tilde D}_{\mu\nu\kappa\lambda}(\partial_\rho
B^{\kappa\lambda} +\partial_\rho H^{\kappa\lambda})
-\partial_\rho C^\rho {\tilde D}_{\mu\nu\kappa\lambda}
(B^{\kappa\lambda}+H^{\kappa\lambda})
$$ $$
+2\gamma_{\rho(\mu}
{\tilde D}^{\rm ghost}_{\nu)\kappa}\partial^\rho H^\kappa
-\partial_\rho\gamma_{\mu\nu}{\tilde D}^{{\rm ghost}\,\rho\kappa}H_\kappa
-\gamma_{\mu\nu}{\tilde D}^{{\rm ghost}\,\rho\kappa}\partial_\rho
H_\kappa]\sigma,
\end{equation}
\begin{equation}
{\hat\delta}_2C_\nu=-\kappa^2{\cal E}_0^2(\partial_\mu C_\nu
{\tilde D}^{{\rm ghost}\,\mu\kappa}H_\kappa+C_\mu
{\tilde D}_{{\rm ghost}\,\nu\kappa}\partial^\mu H^\kappa)\sigma.
\end{equation}
Here, we have
\begin{equation}
H^{\alpha\beta}=-(\partial^{(\alpha}{\bar C}_\rho
\partial^{\beta)}C^\rho+\partial^\rho{\bar C}^{(\alpha}
\partial^{\beta)}C_\rho
+\partial^\rho\partial^{(\beta}{\bar C}^{\alpha)} C_\rho),
\end{equation}
\begin{equation}
H^\rho=\gamma_{\lambda\kappa}\partial^\lambda\partial^\kappa
C^\rho+\partial^\kappa\gamma_{\lambda\kappa}\partial^\lambda
C^\rho
-\partial_\kappa\partial_\lambda\gamma^{\rho\kappa}
C^\lambda- \partial_\kappa\gamma^{\rho\kappa}\partial_\lambda
C^\lambda,
\end{equation}
\begin{equation}
{\bar H}^\rho
=\partial^\lambda{\bar C}^\rho\partial^
\kappa\gamma_{\lambda\kappa}
+\partial^\lambda\partial^\kappa{\bar
C}^\rho\gamma_{\lambda\kappa} +\partial^\rho{\bar
C}^\lambda\partial^\kappa \gamma_{\lambda\kappa}.
\end{equation}
Moreover, ${\tilde D}_{\alpha\beta\mu\nu}$ is the ``stripping'' propagator
for the graviton in the gauge $\alpha=-1$:
\begin{equation}
{\tilde D}_{\alpha\beta\mu\nu}(p)
=\frac{1}{2}(\eta_{\alpha\mu}\eta_{\beta\nu}+\eta_{\alpha\nu}\eta_{\beta\mu}
-\eta_{\alpha\beta}\eta_{\mu\nu}){\cal O}_0(p),
\end{equation}
while the
ghost stripping propagator is given by
\begin{equation}
{\tilde D}^{\rm ghost}_{\mu\nu}(p)=\eta_{\mu\nu}{\cal O}_0(p),
\end{equation}
where
\begin{equation}
{\cal O}_0(p)=\frac{{\cal E}^2_0-1}{p^2}.
\end{equation}
We see that the local propagator can be obtained from the nonlocal
propagator minus the stripping propagator
\begin{equation}
\frac{1}{p^2}=\frac{\exp(p^2/\Lambda_G^2)}{p^2}-{\cal O}_0(p).
\end{equation}
The stripping propagators are used to guarantee that the tree-level
graviton-graviton scattering amplitudes are identical to the local,
point-like tree-level amplitudes, which couple only to physical gravitons.

The graviton propagator in the fixed
de Donder gauge $\alpha=-1$~\cite{Donder} in momentum space is given by
\begin{equation}
D_{\mu\nu\rho\sigma}(p)
=\frac{\eta_{\mu\rho}\eta_{\nu\sigma}+\eta_{\mu\sigma}\eta_{\nu\rho}
-\eta_{\mu\nu}\eta_{\rho\sigma}}{p^2-i\epsilon},
\end{equation}
while the graviton ghost propagator in momentum space is
\begin{equation}
D^{\rm ghost}_{\mu\nu}(p)=\frac{\eta_{\mu\nu}}{p^2-i\epsilon}.
\end{equation}

As in the case of the Yang-Mills sector, the on-shell vertex functions
are unaltered from their local antecedents, while virtual
particles are attached to nonlocal vertex function form factors. This destroys the
gauge invariance of e.g. graviton-graviton scattering and requires an iteratively
defined series of ``stripping'' vertices to ensure the decoupling of all unphysical
modes. Moreover, the local gauge transformations have to be extended to nonlinear,
nonlocal gauge transformations to guarantee the over-all invariance of the
regularized amplitudes. Cornish has derived the primary graviton vertices and the
BRST symmetry relations for the regularized ${\hat W}_{\rm
grav}$~\cite{Cornish,Cornish2}, using the nonlinear, nonlocal extended gauge
transformations suitable for the perturbative gravity equations.

Because we have extended the gauge symmetry to nonlinear,
nonlocal transformations, we must also supplement the
quantization procedure with an invariant measure
\begin{equation}
{\cal M}=\Delta({\bf g}, {\bar C}, C)D[{\bf
g}_{\mu\nu}]D[{\bar C}_\lambda]D[C_\sigma]
\end{equation}
such that $\delta {\cal M}=0$.

As we have demonstrated, the quantum gravity perturbation theory
is invariant under generalized, nonlinear field
representation dependent transformations, and it is
finite to all orders. At the tree graph level all
unphysical polarization states are decoupled and nonlocal effects
will only occur in graviton and graviton-matter loop graphs.
Because the gravitational tree graphs are purely local there is a
well-defined classical GR limit. The finite quantum gravity
theory is well-defined in four real spacetime dimensions or in any higher
D-dimensional spacetime.

We quantize by means of the path integral operation
\begin{equation}
\langle 0\vert T^*(O[{\bf g}])\vert 0\rangle_{\cal E}=\int[D{\bf g}]
\mu[{\bf g}]({\rm gauge\, fixing})
O[{\bf g}]\exp(i\hat W_{\rm grav}[{\bf g}]).
\end{equation}
The quantization is carried out
in the functional formalism by finding a measure factor
$\mu[{\bf g}]$ to make $[D{\bf g}]$ invariant under the
classical symmetry. To ensure a correct gauge fixing scheme, we write
${\hat W}_{\rm grav}[{\bf g}]$ in the BRST invariant form with ghost fields; the
ghost structure arises from exponentiating the Faddeev-Popov
determinant~\cite{Fradkin}.
The algebra of extended gauge symmetries is not expected to close
off-shell, so one needs to introduce higher ghost terms (beyond
the normal ones) into both the action and the BRST
transformation. The BRST action will be regularized directly to
ensure that all the corrections to the measure factor are
included.

\section{Standard Model and Gravitational Vacuum Polarization}

A basic feature of our regularized field theory is that the vertex function
form factors
${\cal F}^k(q^2)$ in momentum space are determined by the nature of the vertex
function. For a SM gauge boson, such as the $W$ or $Z$ boson connected to a standard
model particle and anti-particle, the vertex function form factor in
Euclidean momentum space is
\begin{equation}
{\cal F}^{\rm SM}(q^2)=\exp\biggl(-q^2/2\Lambda^2_{SM}\biggr),
\end{equation}
while for a vertex with a graviton attached to a standard model particle and an
anti-particle, the vertex function form factor will be
\begin{equation}
{\cal F}^G(q^2)=\exp\biggl(-q^2/2\Lambda^2_G\biggr).
\end{equation}
Thus, when two vertices are drawn together to make a loop graph, the energy scale
dependence $\Lambda$ will be determined by the external legs attached to the loop.
If we ignore the weak effects of gravity in SM calculations, then the graviton scale
$\Lambda_G$ can be ignored, as is usually the case in SM calculations.

A calculation of the one-loop
gluon vacuum polarization gives the tensor in D-dimensions~\cite{Kleppe2}:
\begin{equation}
\Pi^{\mu\nu}_{ik}(p)=\frac{e^2}{2^D\pi^{D/2}}f_{ilm}f_{klm}
(p^2\eta^{\mu\nu}-p^\mu p^\nu)\Pi(p^2),
\end{equation}
where $p$ is the gluon momentum and
\begin{equation}
\Pi(p^2)=2\int^{1/2}_0dy\Gamma(2-D/2, yp^2/\Lambda_{\rm SM}^2)
[y(1-y)p^2]^{D/2-2}
$$ $$
\times[2(D-2)y(1-y)-\frac{1}{2}(D-6)].
\end{equation}
We observe that $\Pi^\mu_{ik\,\mu}(0)=0$ a result that is
required by gauge invariance and the fact that the gluon has zero
mass.

The dimensionally regulated gluon vacuum polarization result is
obtained by the replacement
\begin{equation}
\Gamma(2-D/2,yp^2/\Lambda^2_{\rm SM})\rightarrow \Gamma(2-D/2)
\end{equation}
and choosing $p^2 \ll \Lambda^2_{\rm SM}$. In four-dimensions we
get
\begin{equation}
\label{vacpolarization}
\Pi(p^2)=2\int^{1/2}_0
dyE_i(yp^2/\Lambda^2_{\rm SM})[4y(1-y)+1],
\end{equation}
where we have used the relation
\begin{equation}
\Gamma(0,z)\equiv E_i(z)=\int^\infty_zdt\exp(-t)t^{-1}.
\end{equation}

The lowest order
contributions to the graviton self-energy will include
the standard graviton loops, the
ghost field loop contribution and the measure loop contribution. In the
regularized perturbative gravity theory the first order vacuum
polarization tensor $\Pi^{\mu\nu\rho\sigma}$ must satisfy the
Slavnov-Ward identities~\cite{Medrano}:
\begin{equation}
\label{SlavnovWard}
p_\mu p_\rho D^{G\mu\nu\alpha\beta}(p)\Pi_{\alpha\beta\gamma\delta}(p)
D^{G\gamma\delta\rho\sigma}(p)=0.
\end{equation}
By symmetry and Lorentz invariance, the vacuum polarization
tensor must have the form
\begin{equation}
\Pi_{\alpha\beta\gamma\delta}(p)
=\Pi_1(p^2)p^4\eta_{\alpha\beta}\eta_{\gamma\delta}+\Pi_2
(p^2)p^4(\eta_{\alpha\gamma}\eta_{\beta\delta}
+\eta_{\alpha\delta}\eta_{\beta\gamma})
$$ $$
+\Pi_3(p^2)p^2(\eta_{\alpha\beta}p_\gamma
p_\delta+\eta_{\gamma\delta}p_\alpha p_\beta)
+\Pi_4(p^2)p^2(\eta_{\alpha\gamma}p_\beta
p_\delta+\eta_{\alpha\delta}p_\beta p_\gamma
$$ $$
+\eta_{\beta\gamma}p_\alpha p_\delta+\eta_{\beta\delta}p_\alpha
p_\gamma)+\Pi_5(p^2)p_\alpha p_\beta p_\gamma p_\delta.
\end{equation}
The Slavnov-Ward identities impose the restrictions
\begin{equation}
\Pi_2+\Pi_4=0,\quad 4(\Pi_1+\Pi_2-\Pi_3)+\Pi_5=0.
\end{equation}

The basic lowest order graviton self-energy diagram is determined
by~\cite{Leibbrandt,Brown,Donoghue,Duff,Duff2}:
\begin{equation}
\Pi^1_{\mu\nu\rho\sigma}(p)=\frac{1}{2}\kappa^2\int
d^4q {\cal U}_{\mu\nu\alpha\beta\gamma\delta}(p,-q,q-p){\cal F}^G(q^2)
D^{G\alpha\beta\kappa\lambda}(q)
$$ $$
\times D^{G\gamma\delta\tau\xi}(p-q){\cal
U}_{\kappa\lambda\tau\xi\rho\sigma}(q,p-q,-p){\cal F}^G((q-p)^2),
\end{equation}
where ${\cal U}$ is the three-graviton vertex function
\begin{equation}
{\cal U}_{\mu\nu\rho\sigma\delta\tau}(q_1,q_2,q_3) =
-\frac{1}{2}[q_{2(\mu}q_{3\nu)}\biggl(2\eta_{\rho(\delta}\eta_{\tau)\sigma}
-\frac{2}{D-2}\eta_{\mu\nu}\eta_{\delta\tau}\biggr)
$$ $$
+q_{1(\rho}q_{3\sigma)}\biggl(2\eta_{\mu(\delta}\eta_{\tau)\nu}
-\frac{2}{D-2}\eta_{\mu\nu}\eta_{\delta\tau}\biggr)+...],
\end{equation}
and the ellipsis denotes similar contributions.

To this diagram, we must add the ghost particle diagram
contribution $\Pi^2$ and
the measure diagram contribution $\Pi^3$. The dominant finite
contribution to the graviton self-energy will be of the form
\begin{equation}
\label{gravpol}
\Pi_{\mu\nu\rho\sigma}(p)\sim\kappa^2\Lambda_G^4Q_{\mu\nu\rho\sigma}(p^2)
$$ $$
\sim\frac{\Lambda^4_G}
{M^2_{\rm PL}}Q_{\mu\nu\rho\sigma}(p^2),
\end{equation}
where $M_{\rm PL}$ is the reduced Planck mass and $Q(p^2)$ is a
finite remaining part.

For {\it renormalizable} field theories such as quantum
electrodynamics and Yang-Mills theory, we will find that
the loop contributions are controlled by the incomplete
$\Gamma$-function. If we adopt an `effective' quantum gravity
theory expansion in the energy~\cite{Donoghue}, then we would
expect to obtain
\begin{equation}
\Pi_{\mu\nu\rho\sigma}(p)\sim\kappa^2
{\cal G}(\Gamma(2-D/2,p^2/\Lambda_G^2)
Q_{\mu\nu\rho\sigma}(p^2),
\end{equation}
where ${\cal G}$ denotes the functional dependence on the
incomplete $\Gamma$-function. By making the replacement
\begin{equation}
{\cal G}(\Gamma(2-D/2),p^2/\Lambda^2_G)\rightarrow
{\cal G}(\Gamma(2-D/2)),
\end{equation}
we would then obtain the second order graviton loop calculations
using dimensional regularization
~\cite{Leibbrandt,Brown,Donoghue,Duff,Duff2,Veltman}.
The dominant behavior will now be $\ln(\Lambda^2_G/q^2)$ and
not $\Lambda^4_G$. However, in a nonrenormalizable theory
such as quantum gravity, the dimensional
regularization technique may not provide a correct result for the
dominant behavior of the loop integral and we expect the result
to be of order $\Lambda_G^4$. Indeed, it is well known that
dimensional regularization for massless particles removes all
contributions from tadpole graphs and $\delta^4(0)$ contact
terms. On the other hand, our regularized field theory takes into account all leading
order contributions and provides a complete account of all
counterterms. Because all the scattering amplitudes are finite,
then renormalizability is no longer an issue.

The function
\begin{equation}
{{Q_{\mu}}^{\mu\sigma}}_\sigma(p^2)\sim p^4
\end{equation}
as $p^2\rightarrow 0$. Therefore, ${{{\Pi_\mu}^\mu}^\sigma}_\sigma(p)$
vanishes at $p^2=0$, as it should from gauge invariance and for massless gravitons.
We now find that
\begin{equation}
\Pi^G(p^2)\sim \frac{\Lambda^4_G}{M^2_{\rm PL}}.
\end{equation}
Thus, the pure graviton
self-energy is proportional to $\Lambda_G^4$. We shall choose
$\Lambda_G < 10^{-3}$ eV, so that the pure gravitational quantum corrections
to the bare cosmological constant $\lambda_0$ are cut off for energies above $\sim
10^{-3}$ eV.

In contrast to recent models of branes and strings in which the
higher-dimensional compactification scale is lowered to the TeV
range~\cite{Witten}, we retain the classical tree graph GR
gravitation picture and its Newtonian limit. It is perhaps a
radical notion to entertain that quantum gravity becomes weaker
as the energy scale increases towards the Planck scale $\sim
10^{19}$ Gev, but there is, of course, no known experimental
reason why this should not be the case in nature. However, we do
not expect that our weak gravity field expansion is valid at the
Planck scale when $GE^2\sim 1$, although the damping
of the quantum gravity loop graphs could still persist at the
Planck scale. This question remains unresolved until a
nonperturbative solution to quantum gravity is found.

\section{Resolution to the Cosmological Constant Problem}

Zeldovich~\cite{Zeldovich} showed that the zero-point
vacuum fluctuations must have a Lorentz invariant form
\begin{equation}
T_{{\rm vac}\,\mu\nu}=\lambda_{\rm vac}g_{\mu\nu},
\end{equation}
consistent with the equation of state $\rho_{\rm vac}=-p_{\rm
vac}$. Thus, the vacuum within the framework of particle quantum
physics has properties identical to the cosmological constant.
In quantum theory, the second quantization of a classical field
of mass $m$, treated as an ensemble of oscillators each with a
frequency $\omega(k)$, leads to a zero-point energy
$E_0=\sum_k\frac{1}{2}\hbar\omega(k)$. The experimental
confirmation of a zero-point vacuum fluctuation was demonstrated
by the Casimir effect~\cite{Casimir}. A simple evaluation
of the vacuum density obtained from a summation of the zero-point
energy modes gives
\begin{equation}
\rho_{\rm vac}
=\frac{1}{(2\pi)^2}\int_0^{M_c}dkk^2(k^2+m^2)^{1/2}
\sim\frac{M^4_c}{16\pi^2},
\end{equation}
where $M_c$ is the cutoff. Already at the level of the standard
model, we get $\rho_{\rm vac}\sim (10^2\,{\rm GeV})^4$ which is
$55$ orders of magnitude larger than the bound (\ref{vacbound}).
To agree with the experimental bound (\ref{vacbound}), we would
have to invoke a very finely tuned cancellation of $\lambda_{\rm
vac}$ with the `bare' cosmological constant $\lambda_0$, which is
generally conceded to be theoretically unacceptable.

We shall consider initially the basic lowest order vacuum
fluctuation diagram computed from the matrix element in flat Minkowski spacetime
\begin{equation}
M_{(2)}^{(0)}\sim
e^2\int d^4pd^4p'd^4k\delta(k+p-p')\delta(k+p-p')
$$ $$
\times\frac{1}{k^2+m^2}{\rm Tr}\biggl(\frac{i\gamma^\sigma
p_\sigma-m_f}{p^2+m_f^2}\gamma^\mu\frac{i\gamma^\sigma
p'_\sigma-m_f}{p^{'2}+m_f^2}\gamma_\mu\biggl)
$$ $$
\exp\biggl[-\biggl(\frac{p^2
+m_f^2}{2\Lambda^2_{SM}}\biggr)-\biggl(\frac{p'^2+m^2_f}{2\Lambda^2_{SM}}\biggr)
-\biggl(\frac{k^2+m^2}{2\Lambda^2_{SM}}\biggr)\biggr],
\end{equation}
where $e$ is a coupling constant associated with the standard model.
We have considered a closed loop made of a SM fermion of mass $m_f$, an antifermion
of the same mass and an internal SM boson propagator of mass $m$; the scale
$\Lambda_{\rm SM}\sim 10^2-10^3$ GeV. This leads to the result
\begin{equation}
M_{(2)}^{(0)}\sim 16\pi^4g^2\delta^4(a)\int_0^{\infty}dpp^3\int_0^{\infty}dp'p^{'3}
\biggl[\frac{-P^2+p^2+p^{'2}+4m_f^2}{(P+a)(P-a)}\biggr]
$$ $$
\times\frac{1}{(p^2+m_f^2)(p'^2+m_f^2)}\exp\biggl[-\frac{(p^2+p'^2+2m^2_f)}
{2\Lambda^2_{SM}}-\frac{P^2+m^2}{2\Lambda^2_{SM}}\biggr],
\end{equation}
where $P=p-p'$ and $a$ is
an infinitesimal constant which formally regularizes the infinite volume factor
$\delta^4(0)$. We see that $\rho_{\rm vac}\sim M_{(2)}^{(0)} \sim \Lambda_{\rm
SM}^4$. To maintain gauge invariance and unitarity, we must add to this
result the contributions from the ghost diagram and the measure
diagram.

In flat Minkowski spacetime, the sum of all {\it disconnected}
vacuum diagrams $C=\sum_nM^{(0)}_n$ is a constant factor in the
scattering S-matrix $S'=SC$. Since the S-matrix is unitary
$\vert S'\vert^2=1$, then we must conclude that $\vert
C\vert^2=1$, and all the disconnected vacuum graphs can be
ignored. This result is also known to follow from the Wick ordering of the field
operators. However, due to the equivalence principle {\it gravity couples to all
forms of energy}, including the vacuum energy density $\rho_{\rm vac}$, so we can no
longer ignore these virtual quantum fluctuations in the presence of a non-zero
gravitational field.

We can view the cosmological constant as a non-derivative coupling
of the form $\lambda_0\sqrt{-g}$ in the Einstein-Hilbert action
(\ref{EinsteinHilbert}). This classical tree-graph coupling has the effect of
de-stabilizing Minkowski spacetime. Quantum corrections to $\lambda_0$ come from
loops formed from massive SM states, coupled to external graviton lines at
essentially zero momentum. The massive SM states are far off-shell. Experimental
tests of the standard model involving gravitational couplings to the SM states are
very close to being on-shell. Important quantum corrections to $\lambda_0$ are
generated by a huge extrapolation to a region in which gravitons couple to SM
particles which are far off-shell.

Let us now consider the dominant contributions to the vacuum
density arising from the graviton loop corrections. As
explained above, we shall perform the calculations by expanding
about flat spacetime and trust that the results still hold for
an expansion about a curved metric background field, which is
strictly required for a non-zero cosmological constant. Since
the scales involved in the final answer, including the
predicted smallness of the cosmological constant, correspond to a
very small curvature of spacetime, we expect that our
approximation is justified.

We shall adopt a simple model consisting of a massive vector
meson $V_\mu$, which has the standard model
mass $m_V\sim 10^2$ GeV. We have for the vector
field Lagrangian density
\begin{equation}
{\cal L}_V=-\frac{1}{4}(-{\bf g})^{-1/2}{\bf g}^{\mu\nu}
{\bf g}^{\alpha\beta}F_{\mu\alpha}
F_{\nu\beta}+m_V^2V_\mu V^\mu,
\end{equation} where
\begin{equation}
F_{\mu\nu}=\partial_\nu V_\mu-\partial_\mu V_\nu.
\end{equation}
We include in the Lagrangian density an additional piece
$-\frac{1}{2}(\partial_\mu V^\mu)^2$, and the vector field
propagator has the form
\begin{equation}
D^{\rm V}_{\mu\nu}
=\frac{\eta_{\mu\nu}}{p^2+m_V^2-i\epsilon}.
\end{equation}

The graviton-V-V vertex in momentum space is given by
\begin{equation}
{\cal V}_{\alpha\beta\lambda\sigma}(q_1,q_2)
=\eta_{\lambda\sigma}
q_{1(\alpha}q_{2\beta)}-\eta_{\sigma(\beta}q_{1\alpha)}q_{2\lambda}
-\eta_{\lambda(\alpha}q_{1_\sigma}q_{2\beta)}
$$ $$
+\eta_{\sigma(\beta}\eta_{\alpha)\lambda}q_1{\cdot q_2}
-\frac{1}{D-2}\eta_{\alpha\beta}(\eta_{\lambda\sigma}
q_1q_2-q_{1\sigma}q_{2\lambda}),
\end{equation}
where $q_1,q_2$ denote the momenta of the two $Vs$ connected to
the graviton with momentum $p$. We use the notation
$A_{(\alpha}B_{\beta)}=\frac{1}{2}(A_\alpha B_\beta+A_\beta
B_\alpha)$.

The lowest order correction to the graviton vacuum loop will have
the form
\begin{equation}
\label{PolV}
\Pi^{\rm GV}_{\mu\nu\rho\sigma}(p)
=-\kappa^2\int d^4q
{\cal V}_{\mu\nu\lambda\alpha}(p,-q,q-p){\cal F}^G(q^2)
D^{V\,\lambda\delta}(-q) $$ $$
\times{\cal V}_{\rho\sigma\kappa\delta}(-p,p-q,q){\cal F}^G((q-p)^2)
D^{V\,\alpha\kappa}(q-p).
\end{equation}
We obtain
\begin{equation}
\label{Ptensor}
\Pi^{\rm GV}_{\mu\nu\rho\sigma}(p)=-\kappa^2
\int\frac{d^4q}{(q^2+m_V^2)[(q-p)^2+m^2_V]}K_{\mu\nu\rho\sigma}(p,q)
$$ $$ \times\exp\biggl[-(q^2+m^2_V)/2\Lambda^2_G\biggr]
\exp\biggl\{-[(q-p)^2+m^2_V]/2\Lambda^2_G\biggr\},
\end{equation}
where in D-dimensions
\begin{equation}
K_{\mu\nu\rho\sigma}(p,q)=p_\alpha p_\beta p_\rho p_\sigma
+q_\alpha p_\beta p_\rho p_\sigma -q_\alpha q_\beta p_\rho
p_\sigma+(1-D)q_\alpha q_\beta q_\rho p_\sigma
$$ $$
-(1+D)p_\alpha q_\beta q_\rho q_\sigma+(D-1)p_\alpha q_\beta
p_\rho q_\sigma +Dq_\alpha q_\beta q_\rho q_\sigma.
\end{equation}
As usual, we must add to (\ref{PolV}) the contributions from the
fictitious ghost particle diagrams and
the invariant measure diagram.

We observe that from power counting of the momenta in the
integral (\ref{Ptensor}), we obtain
\begin{equation}
\Pi^{\rm GV}_{\mu\nu\rho\sigma}(p)\sim
\kappa^2\Lambda_G^4N_{\mu\nu\rho\sigma}(p^2)
$$ $$
\sim\frac{\Lambda_G^4}{M^2_{\rm PL}}N_{\mu\nu\rho\sigma}(p^2),
\end{equation}
where $N(p^2)$ is a finite remaining part of $\Pi^{\rm GV}(p)$. We
have as $p^2\rightarrow 0$:
\begin{equation}
{{{N_\mu}^\mu}^\sigma}_\sigma(p^2)\sim p^4.
\end{equation}
Thus, ${{{\Pi^{\rm GV}_\mu}^\mu}^\sigma}_\sigma(p)$ vanishes at $p^2=0$,
as it should because of gauge invariance and the massless graviton.

We now have
\begin{equation}
\Pi^{\rm GV}(p^2)\sim \frac{\Lambda_G^4}{M^2_{\rm PL}}.
\end{equation}
If we choose $\Lambda_G\leq 10^{-3}$ eV, then the quantum correction to the
bare cosmological constant $\lambda_0$ is suppressed sufficiently to satisfy the
bound (\ref{lambdabound}), and it is protected from large unstable radiative
corrections. This provides a solution to the cosmological constant problem at the
energy level of the standard model and possible higher energy extensions of the
standard model. The universal fixed gravitational scale $\Lambda_G$ corresponds to
the fundamental length $\ell_G\leq 1$ cm at which virtual gravitational radiative
corrections are cut off.

The vector field vertex form factor, {\it when coupled to SM gauge bosons}, will have
the form
\begin{equation} {\cal F}^{\rm SM}(q^2)
=\exp\biggl[-(q^2+m^2_V)/2\Lambda_{SM}^2\biggr].
\end{equation}
If we choose $\Lambda_{SM} > 1-10$ TeV, then we will reproduce the SM
experimental results, including the running of the SM coupling constants,
and ${\cal F}^{\rm SM}(p^2)$ becomes ${\cal F}^{\rm SM}(0)=1$ on the mass shell
$q^2=-m_V^2$.

We observe that the required suppression of the vacuum diagram
loop contribution to the cosmological constant, associated with
the vacuum energy momentum tensor at lowest order,
demands a low gravitational energy scale $\Lambda_G\leq
10^{-3}$ eV, which controls the quantum gravity loop
contributions. This is essentially because the external
graviton momenta are close to the mass shell, requiring a low
energy scale $\Lambda_G$. This seems at first sight a radical
suggestion that quantum gravity corrections are weak at energies
higher than $\leq 10^{-3}$ eV, but this is clearly not in
contradiction with any known gravitational experiment. Indeed, as
has been stressed in recent work on large higher dimensions,
there is no experimental knowlege of gravitational forces below 1
mm. In fact , we have no experimental knowledge at present about
the strength of graviton radiative corrections. The SM
experimental agreement is achieved for SM
particle states close to the mass shell. However, we expect that
the dominant contributions to the vacuum density arise from
SM states far off the mass shell. In our perturbative
quantum gravity theory, the tree graphs involving gravitons are
identical to the tree graphs in local point graviton perturbation
theory, retaining classical, causal GR and Newtonian gravity. In
particular, {\it we do not decrease the strength of the
classical gravity force.}

In order to solve the severe cosmological constant hierarchy
problem, we have been led to the surprising conclusion that,
in contrast to the conventional folklore, quantum gravity
corrections to the classical GR theory are negligible at
energies above $\leq 10^{-3}$ eV, a result that will continue
to persist if our perturbative calculations can be extrapolated
to near the Planck energy scale $\sim 10^{19}$ GeV. Since the
cosmological constant problem already results in a severe crisis
at the energies of the standard model, our quantum gravity
resolution based on perturbation theory can resolve the
crisis at the standard model energy scale and well beyond this
energy scale.

\section{\bf Conclusions}

The ultraviolet finiteness of perturbative quantum field
theory in four-dimensions is achieved by applying
the nonlocal field theory formalism. The nonlocal quantum loop interactions
reflect the quantum, non-point-like nature of the field theory. Thus, as with string
theories, the point-like nature of particles is `fuzzy' for energies greater than
the scale $\Lambda$. One of the features of superstrings is that they provide a
mathematically consistent theory of quantum gravity, which is ultraviolet finite and
unitary. Our nonlocal theory focuses on the basic mechanism behind string theory's
finite ultraviolet behavior by invoking a suppression of bad vertex behavior at high
energies, without compromising perturbative unitarity and gauge invariance. It
provides a mathematically consistent theory of quantum gravity at the perturbative
level. If we choose $\Lambda_G\leq 10^{-3}$ eV, then quantum radiative corrections to
the classical tree graph gravity theory are perturbatively negligible to all energies
greater than $\Lambda_G$, provided that the perturbative regime is valid.

The important gauge hierarchy problem, associated with the Higgs sector, can also be
resolved in our nonlocal field theory. It can be shown that an exponential damping of
the Higgs self-energy in the Euclidean $p^2$ domain occurs for $p^2 \gg \Lambda_H^2$,
and for a $\Lambda_H$ scale in the electroweak range $\sim 10^2-10^3$ GeV.

A damping of the vacuum polarization loop contributions to the vacuum energy
density-gravity coupling at lowest order can resolve the cosmological constant
hierarchy problem, if the gravity loop scale $\Lambda_G\leq 10^{-3}$ eV, by
suppressing virtual gravitational radiative corrections above the energy scale
$\Lambda_G$. We expect that the SM energy scale $\Lambda_{\rm SM}$ to
be much larger than the electroweak scale $\sim 10^2-10^3$ Gev,
and it could be as large as grand unification theory (GUT) scale
$\sim 10^{16}$ Gev, allowing for possible GUT unification
schemes.

Recently, new supernovae data have strongly indicated a
cosmic acceleration of the present
universe~\cite{Perlmutter}. This has brought the status of
the cosmological constant back into prominence, since one
possible explanation for this acceleration of the expansion of the
universe is that the cosmological constant is non-zero but very
small. We can, of course, accomodate a small non-zero
cosmological constant by choosing carefully the gravity scale
$\Lambda_G$. Indeed, this new observational data can be viewed as
a means of determining the size of $\Lambda_G$.

\vskip 0.2
true in {\bf Acknowledgments}
\vskip 0.2 true in
I thank Michael Clayton, George Leibbrandt and Michael Luke
for helpful discussions. This work was supported by the Natural
Sciences and Engineering Research Council of Canada.
\vskip 0.5
true in


\begin{thebibliography}{100}

\bibitem{Weinberg2} S. Weinberg, Rev. Mod. Phys. {\bf 61}, 1
(1989).

\bibitem{Sahni} V. Sahni and A. Starobinski, astro-ph/9904396.

\bibitem{Witten2} E. Witten, hep-ph/0002297.

\bibitem{Carroll} S. M. Carroll, hep-th/0004075.

\bibitem{Straumann} N. Straumann, astro-ph/0009386.

\bibitem{Rugh} S. E. Rugh and H. Zinkernagel, hep-th/0012253.

\bibitem{Moffat} J. W. Moffat, Phys. Rev. D{\bf 41}, 1177
(1990).

\bibitem{Moffat2} D. Evens, J. W. Moffat, G. Kleppe and R. P. Woodard,
Phys. Rev. D{\bf43}, 49 (1991).

\bibitem{Moffat3} J. W. Moffat and S. M. Robbins, Mod. Phys. Lett.
A{\bf 6}, 1581 (1991).

\bibitem{Woodard} G. Kleppe and R. P. Woodard, Phys. Lett. B{\bf 253},
331 (1991).

\bibitem{Kleppe2} G. Kleppe and R. P. Woodard, Nucl. Phys. B{\bf 388},
81 (1992).

\bibitem{Cornish} N. J. Cornish, Mod. Phys.
Lett. {\bf 7}, 631 (1992).
 
\bibitem{Cornish2} N. J. Cornish, Mod. Phys. Lett. {\bf 7}, 1895
(1992).

\bibitem{Hand} B. Hand, Phys. Lett. B{\bf 275}, 419
(1992).

\bibitem{Woodard2} G. Kleppe and R. P. Woodard, Ann. of  Phys.
{\bf 221}, 106 (1993).

\bibitem{Clayton} M. A. Clayton, L. Demopolous and J. W. Moffat,
Int. J. Mod. Phys. A{\bf 9}, 4549 (1994).

\bibitem{Paris} J. Paris, Nucl. Phys. {\bf B450}, 357 (1995).

\bibitem{Troost} J. Paris and W. Troost, Nucl. Phys. {\bf B482}.
373 (1996).

\bibitem{Joglekar} G. Saini and S. D. Joglekar, Z. Phys. {\bf
C76}, 343 (1997).

\bibitem{Joglekar2} S. D. Joglekar, hep-th/0003104,
hepth/0003077.

\bibitem{Joglekar3} A. Basu and S. D. Joglekar, hep-th/0004128.
 
\bibitem{Moffat4} J. W. Moffat, hep-th/9808091. Talk given at
the XI International Conference on Problems in Quantum Field
Theory, Dubna, Russia, July, 1998. Proceedings published by World
Scientific, Singapore, 1999; J. W. Moffat, Talk given at the IV
Workshop on Quantum Chromodynamics, June, 1998, eds. H. M. Fried
and B. M\"uller. Proceedings published by World Scientific,
Singapore, 1999.

\bibitem{Sundrum} R. Sundrum, JHEP 9907, 001 (1999),
hep-ph/9708329 v2.

\bibitem{Witten} E. Witten, Nucl. Phys. B{\bf 471}, 135 (1996); J. D.
Lykken, Phys. Rev. D{\bf 54}, 3693 (1996); I. Antoniadis, Phys.
Lett. B{\bf 246}, 377 (1990); I. Antoniadis, N. Arkani-Hamed, S.
Dimopolous, and G. Dvali, Phys. Lett. B{\bf 429}, 263 (1998); N.
Arkani-Hamed, S. Dimopoulos, and G. Dvali, Phys. Rev. D{\bf 59},
105002 (1999); K. Dienes, E. Dudas, and T. Gherghetta, Nucl.
Phys. B{\bf 537}, 47 (1999); L. Randall and R. Sundrum, Phys.
Rev. Lett. {\bf 83}, 3370 (1999), hep-th/9905221.

\bibitem{Becchi} C. Becchi, A. Rouet, and R. Stora, Comm.
Math. Phys. {\bf 42}, 127 (1975); I. V. Tyutin, Lebedev Institute preprint
N39 (1975).

\bibitem{Feynman3} R. P. Feynman, Acta Phys.
Pol. {\bf 24}, 697 (1963); Magic Without Magic, edited by J.
Klauder (Freeman, New York, 1972), p. 355; Feynman Lectures on
Gravitation, edited by B. Hatfield, (Addison-Wesley publishing
Co. 1995.)

\bibitem{Veltman} G. 't Hooft and M. Veltman, Ann. Inst. Henri
Poincar\'e, {\bf 30}, 69 (1974).

\bibitem{Van} S. Deser and P. van Nieuwenhuizen, Phys. Rev. Phys.
Rev. {\bf 10}, 401 (1974); Phys. Rev. {\bf 10}, 411 (1974).
 
\bibitem{Sagnotti} M. Goroff and A. Sagnotti, Nucl. Phys. {\bf
B266}, 709 (1986).

\bibitem{Goldberg} J. N. Goldberg, Phys. Rev. {\bf 111}, 315
(1958).

\bibitem{Fradkin} E. S. Fradkin and I. V. Tyutin, Phys. Rev.
{\bf D2}, 2841 (1970).

\bibitem{Donder} T. de Donder, {\it La Grafique Einsteinienne}
(Gauthier-Villars, Paris, 1921); V. A. Fock, {\it Theory of Space, Time and
Gravitation} (Pergamon, New York, 1959).

\bibitem{Medrano}  D. M. Capper and M. R. Medrano, Phys. Rev.
{\bf 9}, 1641 (1974).

\bibitem{Leibbrandt} D. M. Capper, G. Leibbrandt, and M. R.
Medrano, Phys. Rev. {\bf 8}, 4320 (1973).

\bibitem{Brown} M. R. Brown, Nucl. Phys. {\bf B56}, 194 (1973).

\bibitem{Donoghue} J. F. Donoghue, Phys. Rev. {D50}, 3874 (1994).

\bibitem{Duff} D. M. Capper, M. J. Duff, and L. Halpern, Phys.
Rev. {\bf 10}, 461 (1974).

\bibitem{Duff2} M. J. Duff, Phys. Rev. {\bf 9}, 1837 (1974).

\bibitem{Zeldovich} Ya. B. Zeldovich, Pis'ma Zh. Eksp. Teor. Fiz.
{\bf 6}, 883 [JETP Lett. {\bf 6}, 316 (1967)].

\bibitem{Casimir} H. B. G. Casimir, Proc. K. Ned. Akad. Wet. {\bf
51}, 635 (1948).
 
\bibitem{Perlmutter} S. Perlmutter et al., Nature {\bf 391}, 51
(1998); Ap. J. {\bf 517}, 565 (1999); A. Riess, et al., Astron.
Journ. {\bf 117} 207 (1998); B. Schmidt et al., Ap. J. {\bf 507},
46 (1998); P. Garnavich et al., Ap. J. {\bf 509}, 74 (1998).
\end{thebibliography}
\end{document}